\documentclass{nature-hack}

\usepackage{amsmath} 
\usepackage{graphicx}
\usepackage{setspace} 
\usepackage{csquotes}

\usepackage{amssymb}

\usepackage{xfrac}

\usepackage{tikz}

\title{An Experimental-Design Perspective on Population Genetic Variation}

\author{{Andre F. Ribeiro$^{1}$}}

\begin{document}

\maketitle

\begin{affiliations}
\item{ Harvard University, 79 John F. Kennedy St, Cambridge, MA 02138, United States}
\end{affiliations}

\begin{abstract}
We consider the hypothesis that Evolution promotes population-wide genome patterns that, under randomization, ensures the External Validity of adaptations across population members. An adaptation is Externally Valid (EV) if its effect holds under a wide range of population genetic variations. A  prediction following the hypothesis is that pairwise base substitutions in segregating regions must be 'random' as in Erdos-Renyi-Gilbert random graphs, but with edge probabilities derived from Experimental-Design concepts. We demonstrate these probabilities, and consequent mutation rates, in the full-genomes of 2504 humans, 1135 flowering plants, 1170 flies, 453 domestic sheep and 1223 brown rats.\footnote{This research was conducted in 2019.}
\end{abstract}



\section{Introduction}

Whether spontaneous mutations are random is a fundamental and long-standing  question in Evolutionary Theory and central to our understanding of diseases\cite{Kimura:1967aa,Levins:1964aa, Pal:2007aa, Martincorena:2012aa}. As source of genome alterations, patterns in mutations can also help explain the accumulated genetic makeup of organisms. Few patterns have however been found, most of them derived from standard population genetics models, such as decreased mutations in deleterious\cite{Martincorena:2012aa} and homozygous\cite{Amos:2013aa} sites. According to standard models, such as Wright-Fisher, populations can be represented by two quantities, their size $N$ and mutation rate $\theta$. Genetic variation is then seen as mainly the product of genetic drift (which increases genetic variation at random rates) and natural selection (at increased rates). The processes, or population structures, modulating these rates have remained more elusive. There remains little doubt, however, about their importance. Too large drift can make populations lose many advantageous mutants\cite{Mccandlish:2015aa} or demand too large populations for adaptation. Population structures that improve the odds of advantageous mutants\cite{Lieberman:2005aa} be selected, or reduce deleterious\cite{Martincorena:2012aa}, can be crucial to populations' success and survival. The first have been called selection amplifiers and were studied theoretically\cite{Lieberman:2005aa,Chastain:2014aa}, when population structures are represented as graphs. 



A mutation can increase (or decrease) an individual's fitness but not do the same for other population members, possibly sending generations down paths of decreased fitness. Recent research in population genetics has investigated the roles that amplifying population structures\cite{Lieberman:2005aa}, mutation robustness\cite{Draghi:2010aa} and phenotypic diversity\cite{Leinonen:2013aa} can have in the evolutionary process. All these research questions can be seen as a questions about phenotypic generalization across populations. It is still unclear which tools from Discrete Mathematics, the Network Sciences and the Computer Sciences are most useful to formulate them. We propose a framework that employs statistical concepts, builds a link between distinct population structures and the robustness of mutations in them, and, articulates not only the benefits of mutational robustness but also problems and tradeoffs surrounding them. At the core of the approach is the study of the \emph{effects} of mutations in populations and whether such effects generalize across populations, which connects the approach to theories of causation. This alternative interpretation allow us to ask to what extent evolution improves the chances that mutation effects (positive, negative or neutral) generalize to and across genomes in a population. A positive answer suggests that \textbf{increase in the population diversity of complex organisms, brought by increased mutation rates, requires new adaptations to be also valid across these expanded genomic ranges}.


\begin{figure}
\centering
\includegraphics[width=1\linewidth]{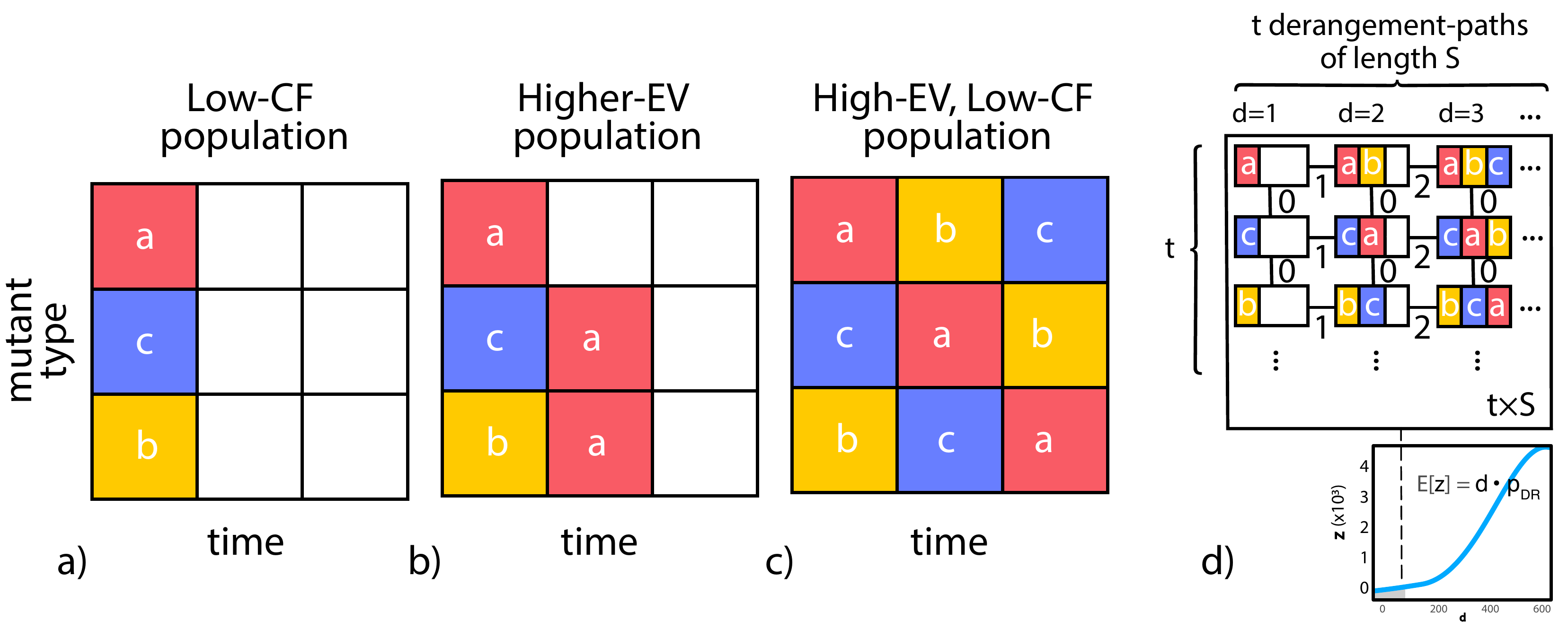}\\
{\small \refstepcounter{figure}\label{fig-pre}\setstretch{0.5}\sffamily\noindent\textbf{Figure \arabic{figure}}\hspace{1em}\textbf{The External Validity and Confounding of distinct combinatorial populations.} Example populations with $S=3$ Single Nucleotide Polymorphisms (SNPs) and distinct mutation rates, members are described by their set of SNPs (colored blocks), grid rows are unique SNP combinations (mutant types) and columns are discrete time intervals, \textbf{(a)} a population with high mutation rate, effects of all individual SNPs can be estimated from pairwise differences in fitness among its members, leading to a population structure with low Confounding ($CF$), \textbf{(b)} population with decreasing mutation rate, leading to a population structure with increased External Validity ($EV$) at position $a$, \textbf{(c)} a Latin-Squares Experimental Design and population with low $CF$ and high $EV$, \textbf{(d, bottom)} number of mutations of size $d< S$ needed to assemble a design with $S$ variables follows a binomial distribution with success probability $p_{\mbox{\tiny DR}}$, \textbf{(top)} we imagine that during selective sweeps Latin-Squares designs grow to the size $t{\approx}S$ of population segregating sites, in the resulting population structure, individuals with the same mutation sizes are mutually disjoint (columns) and intersection sizes increase linearly across nested mutations (rows) - edges are labeled with expected overlap sizes among member types\protect\footnotemark.}
\end{figure}

\footnotetext{some edges were omitted for clarity.}


 
Let population members be described by their set of Single Nucleotide Polymorphisms (SNPs) from a starting, or reference, genome $r$. Fig.(\ref{fig-pre}) illustrates the proposed model when there are $S{=}3$ SNPs - labeled $a$, $b$ and $c$. Grid columns are time intervals, cells are SNPs and rows are unique SNP combinations (population mutant types) at the time. Different mutation rates and population structures can make it easy, or impossible, to distinguish fitness gains brought by these SNPs. If $y_i$ is a fitness measure for member $i$, then fitness differences between any two population members, $y_i-y_j$, can reveal the effect of the set of SNPs the two population members differ in. The leftmost population, Fig.(\ref{fig-pre}a), is a population with high mutation rate. At that instant, all $S$ SNPS are added to the current population $r$.  From all resultant pairwise fitness differences among this population's members, we can distinguish the individual effects of \emph{all} its SNPs. The pairwise genome differences in this population are $\{a,b,c\}$.  We say the population has low Confounding ($CF$). These effects hold certainly, however, for a limited population (i.e., the one with genome $r$). On the middle, Fig.(\ref{fig-pre}b), is a population with decreasing mutation rate. In this case, the baseline population where effects hold expands for the SNP labeled $a$. Effect estimates for $a$ are now possible for populations $r$, $r{+}b$ and $r{+}c$. We say this population has higher External Validity ($EV$). Finally, the rightmost panel shows a Latin-Squares Experimental Design\cite{Mead:2017aa} . The design is associated with populations whose effects have both maximum $EV$ and minimum $CF$. Latin Squares are peculiar in that both their rows and columns are fully pairwise disjoint. Fig.(\ref{fig-pre}d, bottom) shows sample sizes required to reproduce these designs under random sampling with replacement (for an increasing number $d$ of 'variables'). Such concepts will allow us to investigate the \textbf{mutation rates evolution chooses compared to those required by such ideal designs, in distinct evolutionary regimes}. A system with expanding $EV$ must sustain a specific mutation rate through time. Such cases of 'expanding designs', and their resulting population combinatorial patterns, are illustrated in Fig.(\ref{fig-pre}d, top) and discussed in detail below.






\begin{figure}
\includegraphics[width=1\linewidth]{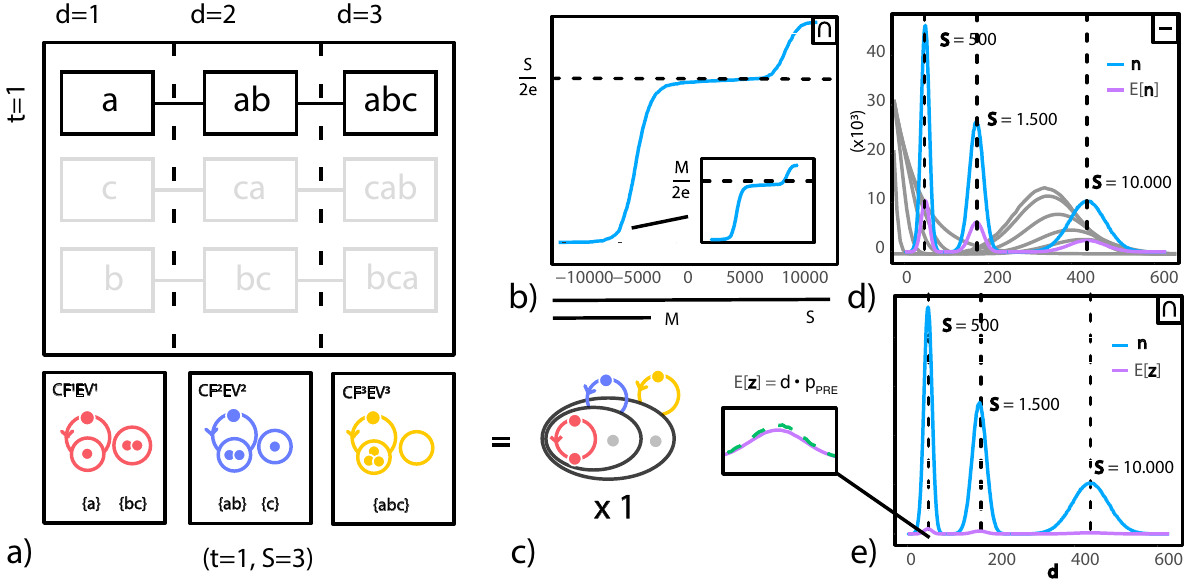}
{\small \refstepcounter{figure}\label{fig-intro}\setstretch{0.5}\sffamily\noindent\textbf{Figure \arabic{figure}}\hspace{1em}\textbf{DNA permutations in populations (a)} an Externally Valid and Unconfounded mutation is described as any single-site mutation leading to the same phenotypic difference under all other possible genetic variations, for $S{=}3$ sites, a mutation of size $d{=}3$ is a \emph{derangement} and $3$-permutation with no fixed-points, a derangement is the permutation with highest level of variation $EV$ (i.e., under which a genetic difference caused a phenotypic), but also highest uncertainty $CF$ (i.e., over which \emph{individual} genomic difference leads to the phenotypic), each mutation from a reference is therefore first labeled $CF^kEV^{d{-}k}$ where $k$ is its $CF$ level and $d{-}k$ its $EV$ level, these labels are illustrated as permutations (bottom, cycles), \emph{fixed-points} are shown as unlooped dots, populations (collections of permutations) can similarly be associated with distinct $CF$ and $EV$ levels, \textbf{(b)} total pairwise overlap size frequency among all individuals in the Human Genome Project\cite{Li:2015aa} for Chromosome 22 (from the mean), $S$ is the number of segregating sites and $M$ a contiguous subset of $S$, illustrating a recursive combinatorial pattern across genomes, \textbf{(c)} a sequence of $S$ derangements of increasing sizes, or \emph{$d$-path}, permutes a site (red) while also permuting all other $S{-}1$ sites, a set of $t$ \emph{simultaneous} $d$-paths leads to a total population variation of $(d{-}k)t$, ${EV}^{(d{-}k)t}$, fitness differences are increasingly generalizable with $t$, we ask how many pairwise differences of each size $d$ need to be observed per site, $\Theta(k,t)$, to guarantee an ${EV}^{(d-k)t}$ and ${CF}^k$ for populations under random sampling? \textbf{(d)} distribution of the expected number of genome differences $E[n]$ (purple) simulated according to Wright-Fisher for $3$ populations with increasing $S$, $n$ are the number of segregating sites per difference size, dotted lines are estimated expected mutation rates\cite{Fu:1995aa}, grey are hypothetical distribution changes with time\cite{Rogers:1992aa}, \textbf{(e)} distribution of population genome intersections $E[z]$ per difference (purple).}\bigskip
\end{figure}





\section*{Genetic Combinatorial Patterns}

A genomic combinatorial pattern is evident from a population's distribution of genetic difference sizes, under equilibrium. This distribution is illustrated in Fig.(\ref{fig-intro}d). The average number of pairwise differences between sample DNA sequences is, as a consequence, connected\cite{Watterson:1975aa,Tajima:1989aa} to mutation rates as 

\begin{subequations}
  \begin{align}
E[\theta]&=4Nu \label{eq-fisher}\\ 
&= \sum_{i=1}^N\sum_{j=i+1}^N \frac{n_{ij}}{\binom{N}{2}}, \label{eq-differences}\\
&= \frac{1}{H_{N{-}1}}S, \qquad (\textrm{equilibrium}) \label{eq-watterson}
\end{align}
\end{subequations}

where $u$ is the neutral mutation rate per site, $n_{ij}$ is the number of Single Nucleotide Polymorphisms (SNPs) between the $i$-th and $j$-th sequences, $S$ is the sample's total number of segregating sites, and $H_{N{-}1}$ is the $(N{-}1)$-th harmonic number. Eq.(\ref{eq-watterson}) only holds at equilibrium. Central to the success of these models is their simplicity. Moments for the number of differences and segregating sites\cite{Fu:1995aa} have been devised. Further insights into the distribution were proposed by considering their out-of-equilibrium shape (e.g., in the widely popular Tajima D\cite{Tajima:1983aa}) and timely evolution\cite{Rogers:1992aa,Tajima:1989ab}. The now standard explanation is that population growth leave characteristic signatures in the distribution\cite{Rogers:1992aa}, as a wave that travels to the right, Fig.(\ref{fig-intro}d, gray). In practice, this is an unsatisfactory explanation however, as natural populations are much more complex, with unknown histories.  Notice that the assumption of random mutations is embedded in Eq.(1) models. Random mutation of DNA sequences with fixed $S$ leads to binomial distributions for their difference sizes. 


Fig.(\ref{fig-intro}b) shows the frequency of pairwise overlap sizes among the genomes of all 2504 individuals in the Human Genome Project\cite{Li:2015aa} (chromosome 22, from the mean, ordered by size). It illustrates the prevalence of intersection sizes $\Big[S(e-1)\Big]/e^2$, due to a random combinatorial pattern that repeats for contiguous segregating regions $M<S$. While Wright-Fisher relate SNP differences and mutation rates, Eq.(\ref{eq-differences}), populations in the model evolve in non-intersecting generations. Patterns in the intersection of populations' genomes have generally been less scrutinized. SNP intersections in populations are equivalent to their frequency there. According to the view here, larger intersections offer the possibility for adaptations' fitness to be observed under different local genomic, or more general, variations. This leads to adaptations that are not only advantageous but also externally valid across the population.  Related to issues encountered here, genetic overlaps have appeared in the study of \textbf{mutation robustness}\cite{Draghi:2010aa,Zheng:2020aa,Fares:2015aa}. There, robustness is often defined as simply the ratio between members with overlapping phenotypes and the total number of phenotypes in the population. When a large set of mutated genomes observe the same phenotype, mutations are said to be robust. It is, however, unclear why robustness should be important and how it relates to adaptability and to mutation rates. We make new empirical predictions by considering the statistical consequences  of increasing population overlapping genomes, exact forms for the distribution of overlaps, and a rationale that explains why and \emph{how} these patterns must change across evolutionary regimes.  

\section*{Model}

The main goal of Theoretical Biology is not to describe, but explain, why (and whether) evolution requires certain distributions or generative processes, such as randomization. Beyond descriptive summaries, the distribution of pairwise differences and intersections remain only loosely connected to the evolutionary process. To that end, we suggest that \textbf{it is better to think of DNA sequence differences in populations as cyclic permutations of DNA subsequences}.  A pair of members with a single nucleotide difference is a permutation of a single site. A pair with $d$ differences permutes $d$ sites. This is illustrated in Fig.(\ref{fig-intro}a). Permutations are depicted as circles (set of site-nucleotides they permute) and arrows (the bijection and exchanged nucleotides). Non-permutated elements are called fixed-points and depicted there as unlooped sites. This alternative characterization is relevant because the amount of variation under which a measure\footnote{measures here are fitness differences $y_i-y_j$ in populations (i.e., the effects of mutations on fitness).}  is taken (represented mathematically here by a permutation) is the central determinant of its validity. A population (and its accumulated pairwise differences) are consequently a collection of permutations, which will determine the external validity and confounding of member adaptations. We will call a permutated site where all other segregating sites have also been permuted in a population the root of a \emph{derangement path}, or, $d$-path. Derangements are permutations with no fixed-points (i.e., that vary fully). Derangement paths can be visualized as a set of nested permutations, Fig.(\ref{fig-intro}c). We will denote $t \in \mathbb{N}^+$ the number of \emph{simultaneous} $d$-paths in a population. 

We use a generating function (g.f.) to formalize this point-of-view and connect patterns of population variation to the External Validity ($EV$) and Confounding ($CF$) of observations there. In the empirical section, we use this g.f. to calculate probability distributions for mutations in populations under different $EV{-}CF$ conditions and random sampling. We consider populations with (1) high $CF$ and low $EV$, (2) low $CF$ and low $EV$, and (3) low $CF$ and high $EV$. We start, however, by enumerating and labeling all individual differences from $r$ (and thus all mutations) with their $CF$ and $EV$ levels. A single mutation of size larger than $1$, $d{>}1$, has the $CF$ level of $d$ and is confounded. We do not know which \emph{individual} site-nucleotide change is responsible for the fitness difference observed between the two individuals. To parse out the effect of \emph{all} SNPs, we need a Latin Square row. The first row of Fig.(\ref{fig-pre}d, top), for example, describes the situation where  we have made enough observations to distinguish the effect of $a$ from $ab$, and of $ab$ from $abc$ - leading, in turn, to possible effect estimates for $\{a,b,c\}$. That's why $CF$ levels are generally associated with the design's columns and $EV$ with rows. A single Latin square row ($t=1$) allow effects to be estimated individually, while multiple rows ($t>1$) for individual effects to be estimated under increasing external conditions. We return to this critical connection later (\emph{Sect.\ref{sect-stat} The Connection Between Genetic Derangements and Fitness}) but this illustrates that individual pairwise differences can be combined to achieve higher $EV$, or lower $CF$, at the cost of larger populations.  In this case, $EV$ and $CF$ become characteristics of populations - and not of individual mutations. 

We formulate next the series describing sample sizes required for populations to achieve distinct $EV$ and $CF$ levels. Generating functions, such as those describing the Galton-Watson process, are common tools in theoretical biology.  The g.f. for the number of mutations per site and $EV$ level is

\begin{equation}\label{eq-god}
\begin{split}
   \Theta(k,t) &=\sum_{d=1}^{\infty} \Bigg[\sum_{i=1}^{t}  \binom{t}{i} \Bigg(\frac{d!}{k!}\sum_{j=1}^{d}  \frac{(-1)^{j}}{j!} \Bigg)^{2-i} \Bigg] \times \mathbf{CF}^{k} \mathbf{EV}^{(d{-}k)t}, 
\end{split}  
  \end{equation}



It is defined over $d = \{0, 1, 2, ...\}$, the infinite nonnegative natural numbers sequence of genomic distances. Its parameters are the number $k$ of fixed sites and/or $t$ of $d$-paths in a population. For example, the g.f. $\Theta(1,1)$ is $1{\times}CF^1EV^1 +2{\times}CF^1EV^2 + 9{\times}CF^1EV^3+...$, which has an increasing $d$ and fixed $k$ and $t$. It indicates that, for a single $d$-path and unconfounded fitness differences, we need $1$ pair with $1$ difference to reach an $EV$ of $1$, $2$ pairs more with $2$ differences to reach an $EV$ of $2$, and $9$ more with $3$ differences for $3$. Each g.f. coefficient is the number of permutations, of each size up to $d$ and $k{-}1$ fixed-points, left after collecting $t$. When $k{=}1$, coefficients are then simply the number of uncollected derangements\footnote{this g.f. is fully derived in the Supporting Material (SM). }. We say that the total amount of variation, $EV$, for each $(d,k,t)$ combination is then $(d{-}k)t$. 

  In particular, a g.f. $\Theta(1,S)$ describes $S$ simultaneous $d$-paths with $EV$ bound by the population's number of segregating sites. It gives the least number of mutations per difference-size one need to observe in order to collect $S$ nested permutations with no fixed-points, Fig.(\ref{fig-intro}c). Sets of $d$-paths lead to distinctive combinatorial patterns in population genomes. We consider testable predictions following from this conceptual framework next.

\begin{figure}
\centering
 \includegraphics[width=1\linewidth]{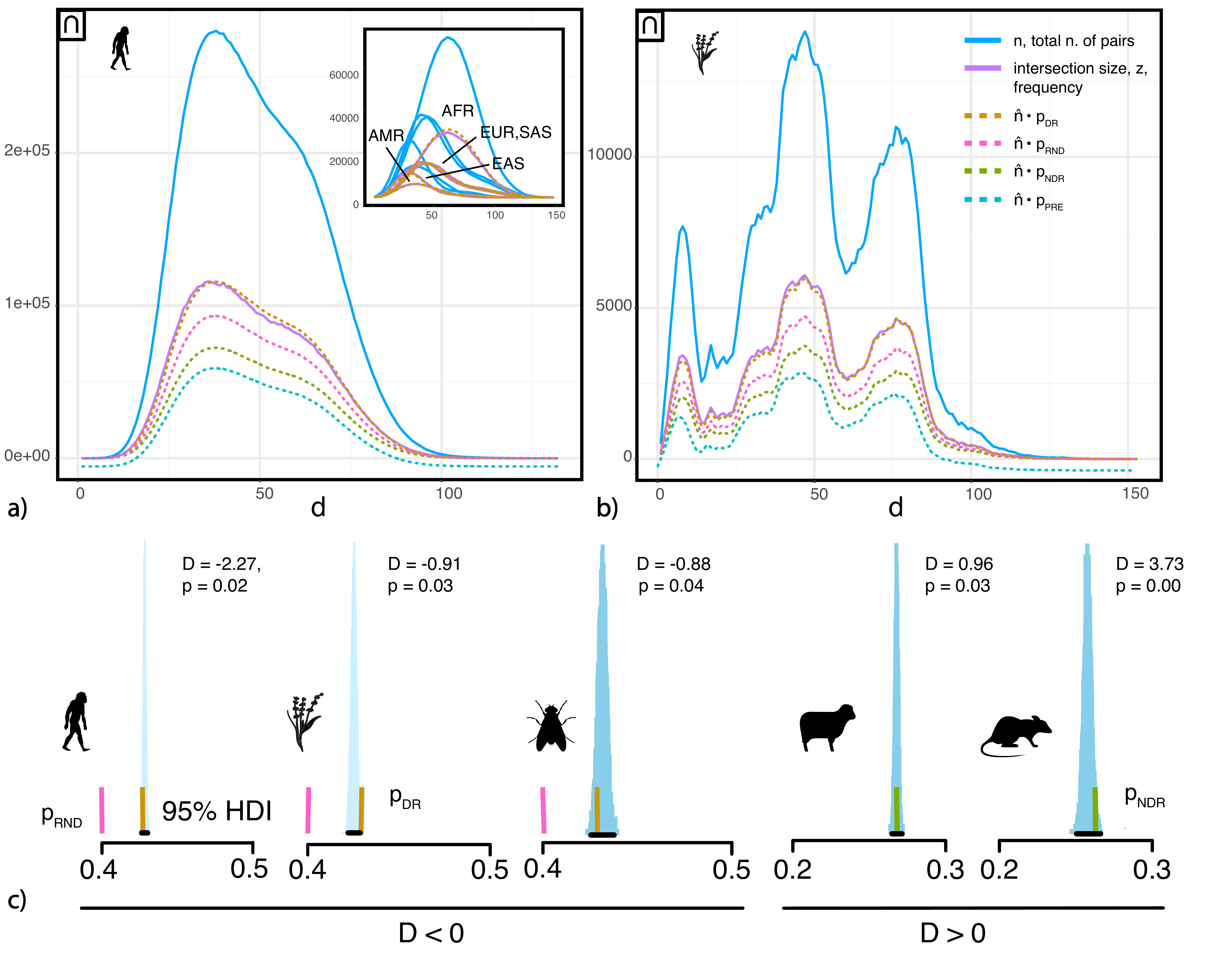}\\
{\refstepcounter{figure}\label{fig-combinatorial}\setstretch{0.5}\sffamily\noindent\textbf{Figure \arabic{figure}}\hspace{1em}\textbf{SNP Intersection Distributions (a)} Chr1:1-2504 first SNPs' Expected intersection-size (purple, solid) by number $d$ of SNPs (pairwise) among the 2504 individuals in the Human Genome Project\cite{Li:2015aa} and all its subpopulations (upper-right panel), according to the proposed, intersections are given by $\hat{n}\cdot p_{cf}$ with $p_{cf}= p_{\mbox{\tiny DR}}$ (brown, dotted) for regions under selective sweeps, as opposed to random alternatives $p_{\mbox{\tiny RND}}$ and $p_{\mbox{\tiny PRE}}$ (Wright-Fisher), and given by $p_{cf} = p_{\mbox{\tiny NDR}}$ for other regimes, $n$ is the total number of segregating site pairs (blue, solid) and $\hat{n}$ the adjusted number susceptible to drift, \textbf{(b)} Chr1:1-1135 region for the 1135 flowering plants (Arabidopsis thaliana) in the 1001 Genomes Project\cite{Alonso-BlancoCarlos20161GRt},  \textbf{(c)} 95\% Credible intervals (HDI) of a Binomial Test for 500 contiguous SNPs with random start positions and length $M=NH_{N-1}$ across the whole genomes of, resp., all human DNA sequences in the 1000 Genomes Project ($N{=}2504$), plants in the 1001 Genomes Project ($N{=}1135$), flies in the Drosophila Reference Panel\cite{Mackay:2012aa} and Nexus\cite{Lack:2016aa} ($N{=}1170$), sheep in the Sheep Genomes Project\cite{Alberto:2018aa} ($N{=}453$) and brown rats in a multi-metropolis study\cite{Combs:2018aa} ($N{=}1223$), success probabilities $p_{\mbox{\tiny DR}}$ (brown), $p_{\mbox{\tiny NDR}}$ (olive) and $p_{\mbox{\tiny RND}}$ (pink) indicated as bars, Tajima's D\cite{Tajima:1983aa} estimates also indicated (top, over the $500M$ concatenated genome sections).}\bigskip
\end{figure}

\section{Experiments}

\section*{Selection Sweeps and Population Genomic Intersections}

Latin Squares have an interesting relation to derangements. Its rows and columns are simultaneous derangements of the considered set $S$ of variables. In particular, they describe the asymptotic case of $t{=}S$. We will first look at horizontal derangements, which we will relate combinatorially to he number $z$ of intersections per $d$ in populations. We later consider vertical derangements, which we will relate to the number of incremental intersections per $d$ in populations. 


 We consider that overlap sizes in populations follow $E[z] = n \cdot p_{cf}$, where $n$ is the total number of pairs over $S$ and $p_{cf}$ is a success probability. \textbf{Erdos-Renyi-Gilbert (ERG) random graph models}\cite{Airoldi:2009aa} provide an alternative interpretation for this model. The central variable in an ERG random graph with $S$ vertices is the probability of an edge, $p$, which is then connected to the expected number of graph edges, $E[n] = n \cdot p$. Expected pairwise population SNP intersections (frequencies) are small in standard model simulations and viral samples. The first is illustrated in Fig.(\ref{fig-intro}e, lower panel) and detailed in the Supporting Material (SM), Fig.(1a,SM), and the second in the SM, Fig.(1b,SM). These expected overlap sizes follow the success probability $p_{cf} = p_{\mbox{\tiny PRE}}$, where $p_{\mbox{\tiny PRE}}=(1-1/e)$. This baseline random combinatorial model is further discussed in the SM. In this letter, we will focus on the combinatorial population patterns of complex and real-world organisms. Other relations to ERG models can also be explored in the future.

 We start with the case of a single $d$-path, $t=1$. The g.f. $\Theta(.,1)$ describes the natural scenario where $EV$ and $CF$ (pairwise differences and intersections) increase together\footnote{for short, we write $\Theta(.,1)$ for the series where both $d$ and $k$ follow the $\mathbb{N}^+{=}\{1,2,3,...\}$ sequence, $k{=}d$.}. This case was illustrated in Fig.(\ref{fig-intro}a). Increase of $EV$ (and pairwise intersections) without increase of $CF$ is also possible, but it requires larger populations. The g.f. $\Theta(1,1)$ indicates that the number of required pairs for each difference size $d$, where $CF$ remains small, is a series $(z_1EV^{1}{+}z_2EV^{2}{+}...)CF^1$. The resulting coefficients $\{z_1,z_2,...\}$ are associated with a discrete probability distribution function. It's easy to derive such distribution, notated $p_{\mbox{\tiny DR}}$, by dividing each coefficient by the number of possible permutations, $d!$, in each $EV$ level. From $\Theta(1,1)$, 
 

\begin{equation}\label{eq-overlap}
\begin{split}
 p_{\mbox{\tiny DR}} &=   \frac{d!}{d!} \sum_{i=1}^{d} \frac{(-1)^i}{i!},\qquad (d\gg k) \\
  &\approx \frac{1}{e}.  
\end{split}  
  \end{equation}



 This probability is effectively independent of $d$. This is due to the known rapid convergence\cite{Hanson:1983aa} of derangements as $d$ increases. It has precision $\left|\frac{\Theta(1,1)}{d!}{-}\frac{1}{e}\right|{<}\frac{1}{(d+1)!}$, which is under 3 decimals for $d$ as low as 4. This g.f. stands in contrast to $\Theta(.,1)$ which is associated with the known probability\cite{Hanson:1983aa} $1/k!e$ that a random permutation of size $d$ have $k$ sites fixed. This probability is different from ${p}_{\mbox{\tiny DR}}$ by a fast-increasing margin. Finally, the pairwise sample requirements prescribed by Eq.(\ref{eq-god}) apply to SNPs that can be lost by drift, which happen with probability $\frac{1}{H_{N-1}}$ in a given sample. The number of pairs $\hat{n} =\frac{1}{1-\frac{1}{H_{N-1}}}n$ is a sample adjustment similar to drop-out adjustments in Experimental studies and made by dividing the recommended sample size by the proportion expected not to be lost. The expected number of SNP intersections is consequently $E[z ] = \hat{n} \cdot p_{cf}$ where $p_{cf} = p_{\mbox{\tiny DR}}$.
 

 Fig.(\ref{fig-combinatorial}) shows observed mutation intersection counts for two example genomic regions, as well as credible intervals for binomial tests over 500 random regions across the whole genomes of 5 species. The figure shows predictions from $\hat{n} \cdot p_{\mbox{\tiny DR}}$ as well as from alternatives success probabilities such as $p_{\mbox{\tiny PRE}}$ and $p_{\mbox{\tiny RND}}$. Empirical observations diverge from these alternatives, as well as from the expected number of random permutations with $k$ fixed-points, $1/k!e$. What is, in fact, observed are numbers very close to the required for populations to increase in $EV$ while keeping $CF$ low. For short, we say we are \emph{deranging the population} in this case. The similarity between predicted and observed distributions is particularly striking after considering the complexity of evolutionary events these populations have undergone. Regions of size $M=NH_{N-1}$ are used, which are relevant to the hypothesis that populations are systematically deranged (where $N$ is the sample size in each case)\footnote{under drift, the mutation rate is $u = 1/4$ in this case, Eq.(\ref{eq-fisher},\ref{eq-watterson}), which is notated $p_{\mbox{\tiny RND}}$ for reference.}.     
 
 Tajima's D estimates across samples, Fig.(\ref{fig-combinatorial}c), indicate that the rate described by $p_{\mbox{\tiny DR}}$ is observed under selective sweeps\cite{Sabeti:2006aa}. The 'expanding design' hypothesis is, however, associated with sustained increase in $t$, which is what we consider next. 
 

\begin{figure}
\includegraphics[width=1\linewidth]{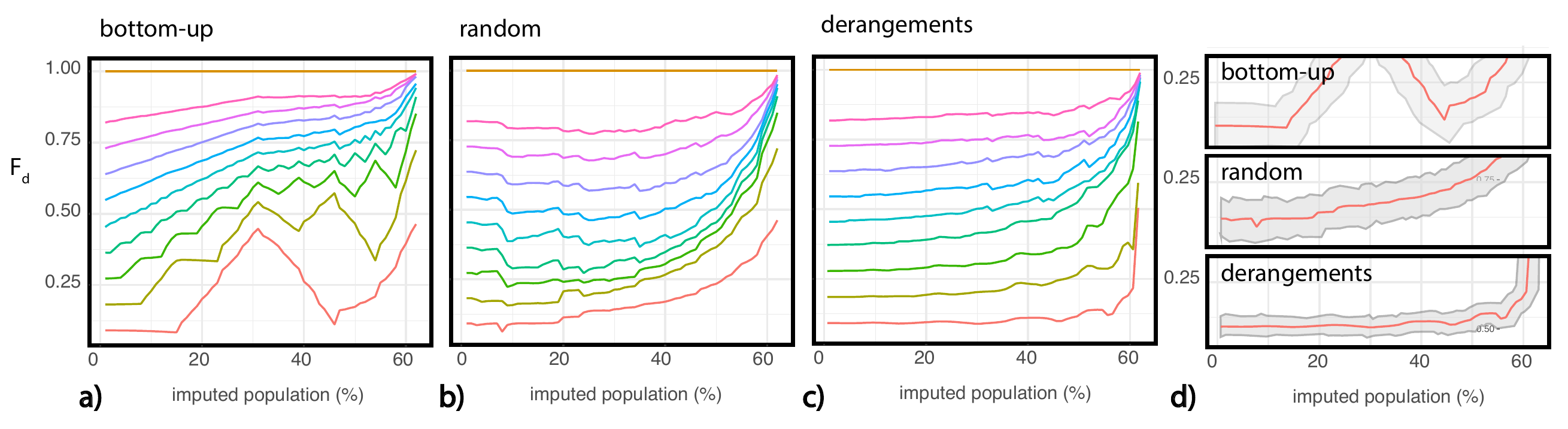}
{\small \refstepcounter{figure}\label{fig-sm}\setstretch{0.5}\sffamily\noindent\textbf{Figure \arabic{figure}}\hspace{1em}\textbf{Simulation} Simple illustration of the relationship between EV and fitness, a genome with 10 sites leads to 3M ($10!$) hierarchical hypotheses, considering 2 nucleotides ($\{A,T\}$) and a population of size 1000 where each individual-site's nucleotide is drawn uniformly, sites with $A$ have an hypothetic effect of $1$ on fitness and their combinations have the mean effect of all their subsets' effects plus a non-additive effect drawn from $[-0.01,+0.01]$ uniformly, there are 10 $EV$ levels (colors), $0 < d \leq 10$, orange ($d{=}10$) is the all-$A$ genome subpopulation and red ($d{=}1$) is the single-$A$ subpopulation, at instant $i=0$ (horizontal axis) the graph shows the outcome $F_{d}$ of an hierarchical ANOVA\cite{Yang:1998aa,Goudet:2005aa} with a complete model specification (where $F_{d}$ is the F-statistic between subpopulation $d$ and $d{+}1$),  at each instant $i>0$ the estimation is repeated after the population is imputed by $10$ genomes, first, in a \textbf{(a)} breadth-first ordering from the bottom, leading to a simpler population model but with increased unexplained variance, the increased effect inaccuracy and decreased $EV$ within subpopulations thus correspond to vertical displacements (lines), \textbf{(b)} imputation of randomly selected genomes and \textbf{(c)} of randomly selected $d$-paths, effects remain accurate across subpopulations up to when 15\% genomes are imputed, or unobserved, in the first case (a), and over 50\% in the last (c), \textbf{(d)} repeated results (mean and bounds) for 1000 simulations and $d=1$.}\bigskip
\end{figure}

\section*{The Connection Between Genetic Derangements and Fitness}\label{sect-stat}


Each difference of size $d$ is a single permutation, and, in order to increase the external validity of fitness differences, we are required to fully permute pairs' fixed-points (i.e., intersections) across the population, Fig.(\ref{fig-intro}c). We called sets of sites with this sequential property derangement paths, or, $d$-paths. Most students are introduced to derangements thorough the Hat-Check problem:

\begin{displayquote}
'$d$ mathematicians drop off their hats at the opera, after the show, their hats are returned at random, what is the chance that no one gets their own hat back?' 
\end{displayquote}

Here, we considered the probability that pairs are deranged in a population corresponds to the requirement that, in expectation, their common sites are varied. The central concept behind Eq.(\ref{eq-god}) is that the $EV$ of adaptation in populations depends on how many derangements it contains.          The minimum number of pairs that guarantees at least one derangement is approximately $d/e$ for a difference of size $d$. To illustrate the effect of increasing derangements on $EV$, consider a particular site, such as the one labeled $a$ in Fig.(\ref{fig-pre}c), and all non-overlapping paths of increasing lengths ending at $a$ in the population. Fig.(\ref{fig-pre}c) shows paths $\{\varnothing,c,bc\}$. The sites preceding $a$ are distinct combinations $x$ of other sites. Let $f_x(a)$ be the effect (fitness difference) of $a$, after $x$. This is associated with a population pair with intersection $x$ and difference $a$. We define the $EV$ of a population as the variance of values $f_x(a)$, $\sigma^2[f_x(a)]$, where each $x=\{x_1,x_2,...,x_d\}$ are $d$ sequential derangements of sizes $\{1,2,...,d\}$. Remember that a permutation of the set of segregating sites is a 1-1 mapping of the set onto itself and a derangement is a permutation with no fixed-points, such that no element appears in its original position. The more derangements we observe, the more variations we measured a site's effect under, and more confident we are the effect will generalize to distinct populations. 

This points to a curious connection to Game Theory. As an abstract measure, the Shapley value\cite{Peters:2015aa}$^{\, \textrm{(Chap. 17)}}$ of an element $a$ of a player-set with size $d$ is the mean payoff difference resulted from the addition of $a$ to all its possible subsets. That is, draw a permutation $\sigma \in \Pi(d)$, each with probability $1/d!$. Then let players enter a room one-by-one in the order $\sigma$ and give each player the marginal contribution created by him. Then player $a$'s payoff is her expected payoff according to this repeated random procedure (or its asymptotic 'effect' given the player-set). To the problem at hand, causal effects are related to the expected Shapley value of a variable, while $EV$ to their variance across the number of observed contexts\footnote{notice that in this letter we simply equate $EV$ with the level of variation required to estimate it, when using the same concepts to estimate causal effects\cite{ribeiro-biobank} this more specific definition becomes useful.}. This also implies that total variation increases with each $d$-path, for all current simultaneous $d$-paths, which makes it natural to employ a coalitional, or cooperative, concept to understand how joint sample sizes scale.   

Fig.(\ref{fig-sm}b-d) describes a simple simulation (see caption) demonstrating these concepts in a population with 1000 individuals and 10 segregating sites. It illustrates that maintaining sequential derangements in a population leads to significantly reduced biases in fitness comparisons when compared to, for example, random and bottom-up genome imputations. In the \emph{Supporting Material} and in a concurrent article\cite{ribeiro-biobank}, we demonstrate that Machine Learning methods and Causal Effect Estimators are subject to these same general constraints. The importance of interacting and dependent effects among SNPs have been demonstrated conclusively by several Genome-Wide Association Studies\cite{Torkamani:2018aa}. Associated with this new outlook on genetics should be Evolutionary mechanisms and strategies that can deal with these more complex statistical landscapes.

\section*{A Limit on Sequential Population Intersections}

We next consider the case of $t{>}1$ and, in particular, $t \approx D_d$ where $D_d$ is the total number of derangements of size $d$. This is the asymptotic case of a large number of $d$-paths. The g.f. $\Theta(1,t)$ indicates that the required number of pairs is $(n_1 EV^{1} {+}\allowbreak n_2EV^{2t}{+}\allowbreak n_3EV^{3t} {+}\allowbreak ... )CF^{1}$. It describes the number of segregating site pairs necessary to collect \emph{all} $D_d$ derangements available, for each difference of size $d$. The resulting coefficients $\{n_1,n_2,...\}$ are associated with the discrete probability distribution of the limiting number of derangements. In the SM, we demonstrate that $\Theta(1,t)$ can be formulated as the product of $D_d$, the number of derangements of a difference of size $d$, and the probability that all derangements have been collected,  

\begin{equation}\label{eq-diff}
\begin{split}
  p_{\mbox{\tiny NDR}} &= \Big[ {d!}\sum_{j=1}^{d}  \frac{(-1)^{j}}{j!} \Big] \Big(1- \frac{1}{e} \Big) \times \frac{1}{d!}, \qquad (t \approx D_d)\\
    &= p_{\mbox{\tiny DR}} \cdot  \Big(1- \frac{1}{e} \Big), \\  
   &\approx \frac{1}{e} \Big( 1 - \frac{1}{e} \Big), \\
   &= \frac{e-1}{e^2},
\end{split}  
  \end{equation}

Each $d$-path is a unique path of length $d$ with distinct derangements for each $EV$ level, Fig.(\ref{fig-intro}c). All $d$-paths have thus the same length but share no fixed-points. The more such paths we collect, the higher the amount of variation in the population. The asymptotic full set of $t \approx D_d$ derangements is associated with a \textbf{Latin Squares Experimental Design} (and a $t {\times} t$ array filled with $t$ different sites, each occurring once in each row and once in each column). As we collect more derangements, the problem of finding new ones becomes increasingly difficult. This is reflected in the difference between $\Theta(1,1)$ and $\Theta(1,t)$, and, consequently, in the two previous success probabilities for drawing derangements, $p_{\mbox{\tiny DR}}$ and $p_{\mbox{\tiny NDR}}$. In fact, we can think of the the derangement process as a transition between these success probabilities. The first describes systems at the begining, when is easier to find derangements. The latter describe systems at the end.

\begin{figure}
\centering
 \includegraphics[width=1\linewidth]{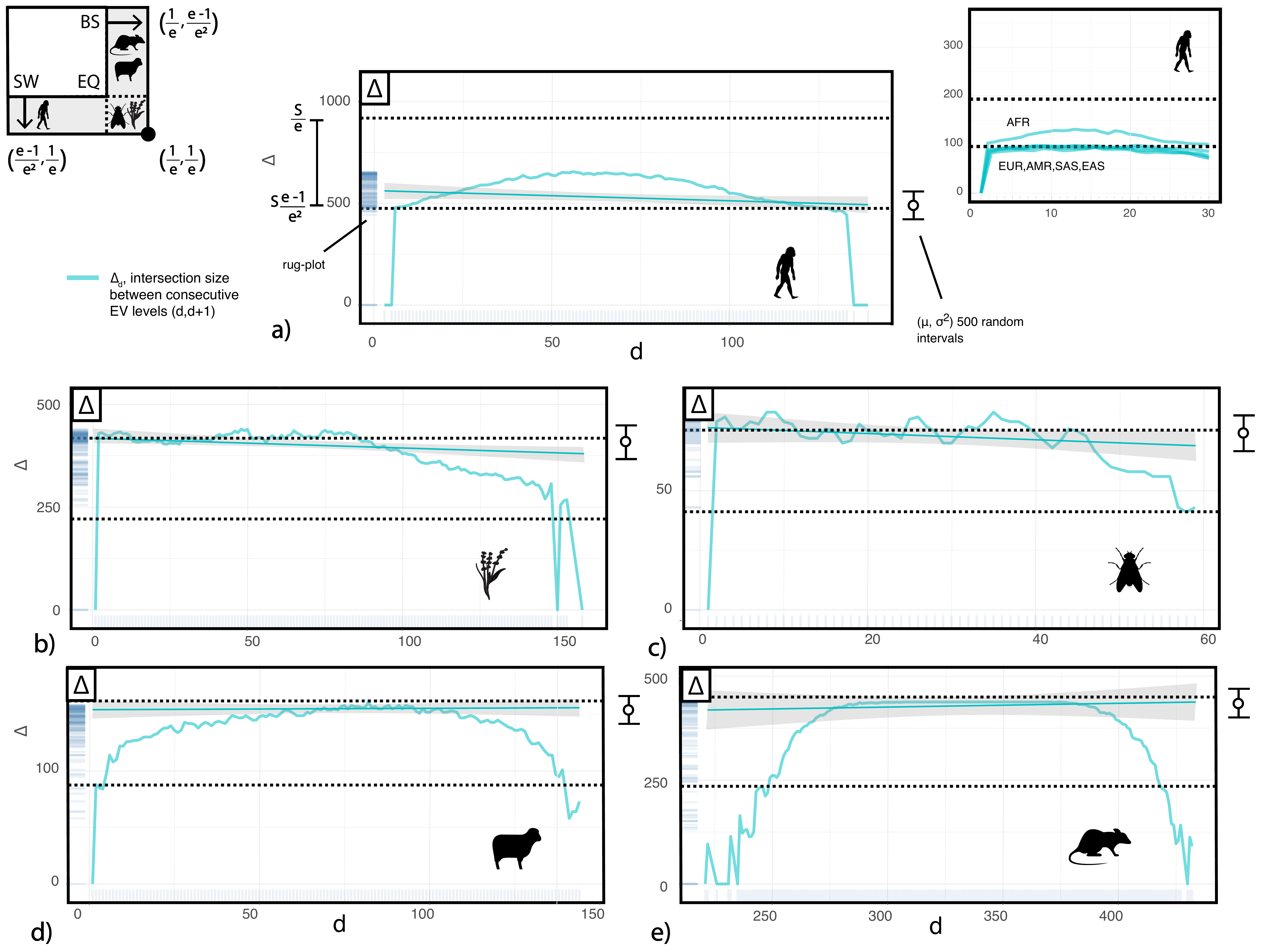}\\
{\refstepcounter{figure}\label{fig-steps}\setstretch{0.5}\sffamily\noindent\textbf{Figure \arabic{figure}}\hspace{1em}\textbf{SNP Sequential Intersections (a)} Chr1:1-2504 SNP intersection sizes $\Delta$ between sequences with increasing difference sizes in human DNA sequences (green, $N{=}2504$) and its subpopulations (upper-right), \textbf{(b)} plants ($N{=}1135$), \textbf{(c)} flies ($N{=}1170$), \textbf{(d)} sheep ($N{=}453$) and \textbf{(e)} brown rats ($N{=}1223$), also shown are the mean and variance of  $\Delta$ across 500 $M$-sized consecutive SNPs with random starting positions (right), according to the proposed, intersections between $d$ are constant and bound by $\Delta =\hat{n}\cdot p_{ev}$ with $p_{ev} \in \{ 1/e, (e-1)/e^2 \}$ (dotted lines), also indicated are the rug-plots for these observations (blue, left), the latter regime is associated with the amount of variation required by Latin Squares Experimental Designs, \textbf{(upper-left panel)} possible expansion directions for these designs and their associated probabilities $(p_{cf},p_{ev})$ (EQ = equilibrium, SW = selection sweep, BS = balancing selection).}\bigskip
\end{figure}





To consider $d$-paths across species, we examine intersections \emph{between} consecutive $EV$ levels, where $\Delta$ for $d$ is the increase in intersecting SNP counts between genomes with differences $d$ and $d{+}1$. This combinatorial pattern was illustrated in Fig.(\ref{fig-pre}d, top) when discussing the 'expanding design' hypothesis. When $t>1$, each site and SNP belong to multiple $d$-paths. Sequential intersections indicate $t$ and are expected to accumulate linearly per $EV$ level. Due to the symmetry in Latin Squares, we can consider similar models vertically and horizontally.  We associate $t$ with a distribution $p_{ev}$ and expected incremental intersections with $E[\Delta ] = S\cdot p_{ev}$. The case of large $t$ is described by $p_{ev} = p_{\mbox{\tiny NDR}}$, and, of small $p_{ev} = p_{\mbox{\tiny DR}}$. This leads to two further predictions: that intersection increments are constant and take one of two values, $p_{ev} \in [ p_{\mbox{\tiny DR}}, p_{\mbox{\tiny NDR}} ]$. Furthermore, constant intersection increments naturally lead to 'nested' subpopulations like the one illustrated in Fig.(\ref{fig-intro}c) (for $t{=}1$).

Fig.(\ref{fig-steps}) plots empirical $\Delta$ across species (green). It is hard not to notice how rapidly these intersection sizes plateau, with a mostly constant $\Delta$. The plateau values also coincide with the previous limits (dotted lines).  Human populations, in particular, have increments described by the lower limit which describe systems with large $t$. A possible explanation might be the simultaneous expansion of environmental variation and population (fulfilling both the need and theoretic requirements for fast increases in $EV$)\cite{Henn:2012aa} . This is supported by the distinctive similarity of $\Delta$ across all subpopulations, except the African (upper-right panel). Previous results are summarized in the upper-left panel in Fig.(\ref{fig-steps}). It depicts the two possible expansion directions (horizontal and vertical) for Latin Square designs, described, in turn, by probabilities $(p_{cf},p_{ev})$. Equilibrium (EQ, black dot) is characterized by the simultaneous deranging of populations in both directions, $(p_{cf},p_{ev}) = ( \sfrac{1}{e},\sfrac{1}{e})$. Both derangement types are 'easily' sampled and have not accumulated in populations. Selection sweeps (SW) are seen as an acceleration of this same process in one direction. Systems under SW are described by $(p_{cf},p_{ev}) = ( \sfrac{1}{e},\sfrac{e{-}1}{e^2})$. They have the proposed minimal number of horizontal derangements in populations, but an increasing $t$ (number of $d$-paths). These are the two conditions formulated for increases in $EV$ with small $CF$. The reverse is observed for populations under balancing selection.




\section*{Conclusion}





Here we considered that evolution is not immune to statistical issues familiar to any researcher, and, that it may make strategic use of randomization in response. We looked at mutations, and population genome variation, as DNA permutations and demonstrated that this interpretation and ensuing predictions, Eq.(\ref{eq-god}), lead to distinctive combinatorial patterns across the genomes of 5 genetic workhorse species. Each of these patterns places direct constrains on the ranges of observable Eukaryote populations. 



The biological puzzle of mutations calls for explanations across different levels, such as its underlying evolutionary strategies\cite{Martincorena:2012aa}, resulting population patterns\cite{Rogers:1992aa,Lieberman:2005aa} and biological mechanisms\cite{Ebersberger:2002aa,Amos:2013aa}. Many other consequences of the $EV{-}CF$ duality need to be studied, such as its genome, temporal and phylogenic limits. Statistically grounded concepts like External Validity and Confounding (in contrast to more abstract others like mutational robustness\cite{Draghi:2010aa,Zheng:2020aa,Fares:2015aa} and graph topologies\cite{Lieberman:2005aa,Chastain:2014aa}) offer rich theoretical possibilities and testable hypotheses when relating population structures to adaptability. These concepts could become useful across other evolutionary levels as well. Not just adaptations, as studied, but successful organisms themselves evolve to occupy increasingly varied environments, while optimizing their use of the genetic code.


\section*{References}
\bibliography{bib}

\end{document}